\documentclass[10pt,final,journal,twoside]{IEEEtran}
\usepackage[mathscr]{eucal}
\usepackage{ifpdf}
\usepackage{cite}
\usepackage{graphicx}
\usepackage[cmex10]{amsmath}
\usepackage{amssymb}
\usepackage{algorithmic}
\usepackage{units}
\usepackage{setspace}
\usepackage{algorithm}
\usepackage{array}
\usepackage[tight,footnotesize]{subfigure}
\usepackage{amsthm}
\usepackage{slashbox}
\usepackage{multirow}
\usepackage{enumerate}
\usepackage{color}
\usepackage{amsmath}
\newcounter{assume}

\newtheorem{theorem}{Theorem}

\newtheorem{proposition}{Proposition}

\begin{document}
\title{Distributed Cell Association for Energy Harvesting IoT Devices in Dense Small Cell Networks: A Mean-Field 
Multi-Armed Bandit Approach}
%
\author{
\IEEEauthorblockN{Setareh Maghsudi and Ekram Hossain, \textit{Fellow, IEEE}\\}
\thanks{The authors are with the Department of Electrical and Computer Engineering, University of Manitoba, Winnipeg, MB, Canada (e-mails: \{setareh.maghsudi, ekram.hossain\}@umanitoba.ca). The work was supported by a CRD grant from the Natural Sciences and Engineering Research Council of Canada (NSERC).}
}
\maketitle
%
\begin{abstract}
The emerging Internet of Things (IoT)-driven ultra-dense small cell networks (UD-SCNs) will need to combat a variety of challenges. On one hand, massive number of devices sharing the limited wireless resources will render centralized control mechanisms infeasible due to the excessive cost of information acquisition and computations. On the other hand, to reduce energy consumption from fixed power grid and/or battery, many IoT devices may need to depend on the energy harvested from the ambient environment (e.g., from RF transmissions, environmental sources). However, due to the opportunistic nature of energy harvesting, this will introduce uncertainty in the network operation. In this article, we study the distributed cell association problem for energy harvesting IoT devices in UD-SCNs. After reviewing the state-of-the-art research on the cell association problem in small cell networks, we outline the major challenges for distributed cell association in IoT-driven UD-SCNs where the IoT devices will need to perform cell association in a distributed manner in presence of uncertainty (e.g., limited knowledge on channel/network) and limited computational capabilities. To this end, we propose an approach based on mean-field multi-armed bandit games to solve the uplink cell association problem for energy harvesting IoT devices in a UD-SCN. This approach is particularly suitable to analyze large multi-agent systems under uncertainty and lack of information. We provide some theoretical results as well as preliminary performance evaluation results for the proposed approach.
\end{abstract}
{\em Keywords}: Internet of Things (IoT), ultra-dense small cell networks (UD-SCNs), energy harvesting, distributed control, cell association, multi-armed bandits, mean-field analysis.
\section{Introduction}
\label{sec:Introduction}
Although a significant growth in mobile data traffic within the next decade is envisioned, traditional cellular networks suffer from numerous shortcomings such as limited uplink capacity, poor cell-edge coverage, and heavy loads at macro base stations (MBSs), all affecting the users' experience adversely. Thus, implementing low-cost, low-power small base stations (SBSs) in order to offload the MBSs traffic and increasing the uplink capacity is foreseen as a promising solution to deliver the expected services of next generation wireless networks, i.e., to sufficiently support human-centric 
(human-to-human [H2H]) communications as well as machine-type (machine-to-machine [M2M]) interactions that exclude human interventions. In general, every small cell is expected to serve up to few hundreds devices; thus, a dense deployment of SBSs is necessary to technically support the massive growth of smart devices in wireless networks, giving rise to paradigms such as the Internet of Things (IoT). As an immediate consequence, it becomes imperative to search for new mathematical tools that are suitable for handling a variety of problems raised by a dense deployment of SBSs as well as the large number of end-users. As specific examples, network management becomes very challenging due to the following reasons. On one hand, centralized methods are not suitable since they require excessive amount of information and impose large computational cost. On the other hand, traditional distributed control mechanisms, such as those adapted from game theory, either yield slow and costly convergence or require heavy data flows between neighboring nodes. Clearly, both problems are aggravated when the number of network nodes become too large.

Another important concern in future ultra-dense small cell networks (UD-SCNs) is to obtain the required energy, both at SBSs- and at end-users. From one side, SBSs are irregularly deployed, so that not all of them can be connected to a power grid. From the other side, the limited battery lifetime of human-centric devices or sensor nodes, thus the requirement of frequent recharge, is an immediate challenge. To date, two main solutions are foreseen to mitigate energy problems: i) wireless energy transfer and ii) energy harvesting. While the former requires a dedicated RF power source, in the latter, the energy is obtained from the ambient (non-dedicated) environment, for instance, solar energy and wind power. This might also include the ambient RF energy from sources such as television or interferences that can be harvested freely from the ambient. As ambient energy harvesting is opportunistic and random in nature, it counts as a source of uncertainty. Facing this uncertainty, any decision making turns to be more challenging for energy-harvesting nodes. In contrast, dedicated wireless charging is more predictable and thus, more reliable; nonetheless, it is prone to storage and transfer loss, so that the efficiency can be achieved only through directional transmission. This necessitates the acquisition of accurate channel state information (CSI), which is however not an easy task specifically when a large number of devices must be powered.  

In this article, we focus on the problem of distributed control of IoT-driven UD-SCNs, more specifically, on the distributed uplink cell association of energy harvesting IoT devices in these networks. As described before, distributed control in such networks is a twofold challenge: first due to the dense deployment of SBSs as well as the large number of devices (e.g., sensors), and second due to the uncertainty introduced by energy harvesting. We argue that in energy-harvesting UD-SCNs, a variety of optimization problems can be formulated as distributed decision making problems in a multi-agent system, where each solution corresponds to an outcome of the interactions of a large number of agents under uncertainty. While confining our attention to the uplink cell association problem, we review the state-of-the-art of research on this problem and outline the open challenges and future research directions. Then, we review a recently-developed mathematical tool, namely mean-field multi-armed bandit games, that can be used to model and analyze the distributed control problems (for instance the cell association problem, among many others) in IoT-driven UD-SCNs. Moreover, we discuss its limitations and possible improvements. To this end, we develop a mean-field multi-armed bandit model for the uplink cell association problem in a UD-SCN, where a large number of IoT devices (human-driven or machines) intend to select an appropriate SBS to become connected to the Internet. We solve the formulated problem and provide some preliminary results.
\section{IoT-Driven UD-SCNs: Challenges, State-of-the-Art, and Open Issues}
\label{sec:StFu}
\subsection{Challenges}
In UD-SCNs, due to irregular and very dense deployment, not all small cells can be connected to a power grid or power beacon. As SBSs have small radio foot print and provide services to nearby users using low transmission power, a feasible solution would be to gain the required energy through ambient energy harvesting. This is usually realized by placing energy harvesting units near small cells, to harvest and convert the ambient energy, such as wind or solar energy. In 
IoT-driven UD-SCNs, not only the SBSs, but also other network entities including machines as well as wearable devices might harvest energy. In particular, ambient RF energy harvesting 
\cite{Kamalinejad15:WEH} has been shown to be a feasible solution for generating small amounts of energy, to be used by network entities such as sensors or electronic wearable devices, in order to reduce the dependency on batteries and other fixed energy resources so that frequent recharge becomes unnecessary. Harvest-store-use (HSU) and harvest-use-store (HUS) are two well-known energy harvesting strategies with the latter being known to be more efficient. In the HSU mode, the stored energy is susceptible to storage loss, whereas in the HUS mode, the energy is utilized right after being harvested and the remainder, if any, is stored in the device \cite{Yuan15:OHUS}. Despite gaining energy at low cost, ambient energy harvesting is opportunistic and depends on non-deterministic factors to a large extent, an inherent characteristic that introduces uncertainty to the system. 
   
One of the main goals of deploying SCNs is to reduce the MBS traffic by offloading some users' traffic to SBSs. To this end, every user equipment becomes associated with some SBS, with the problem being named as \textit{cell association} or \textit{user association}. Although user association is a fundamental problem also in traditional cellular networks, it is significantly more challenging in energy harvesting UD-SCNs, since every user might be located within the coverage area of a set of SBSs, determined by the random amount of harvested energy at SBSs and/or the user itself, as well as random channel gains. Moreover, due to small coverage radius, for mobile users, handover has to be performed rather frequently. In addition, due to the absence of a dedicated central controller, cell association should be preferably performed in a distributed manner. In this paper we focus on cell association problem as an example of general control problem in UD-SCNs. 
\subsection{State-of-the-Art}
In the existing literature, cell association in SCNs is performed in conjunction with a variety of objectives, including interference mitigation, capacity maximization and energy efficiency as most important examples. Nonetheless, only a few of the research works address such problems under the assumption that SBS and/or users rely on ambient energy harvesting as power resource. In the following we discuss state-of-the-art.
\subsubsection{Interference mitigation}
In a dense SCN, in order to retain spectral efficiency, the spectrum shall be reused even by closely located small cells. Thus, although by compacting the network with a large number of SBSs any user might enjoy receiving a strong signal from the SBS to which it is associated, the interference from neighboring cells increases as well, especially for cell-edge users. Therefore, not only intra-cell interference, but also inter-cell interferences should be mitigated in order to reach the expected capacity of UD-SCNs and high  satisfaction level for users. Spatial interference cancellation through cell association is investigated in \cite{Li:SIC}. Stochastic geometry is used to model and solve the cell association problem. Similarly, interference mitigation through power control and cell association (which is also referred to as user association) is investigated in \cite{Chiang:JPC}. The problem is formulated as an integer program and a heuristic algorithm is proposed to solve it. None of these works, however, considers energy harvesting along with interference mitigation.
\subsubsection{Capacity maximization}
\label{subsubsec:ThMax}
Similar to any other network architecture, capacity maximization is a fundamental problem also in SCNs, giving rise to a large body of literature. In \cite{Maghsudi15:DUA}, the authors investigate a distributed user association problem in the downlink of a SCN where SBSs obtain the required energy through ambient energy harvesting. The SCN is modeled as a competitive market with uncertainty, where SBSs and users are represented as consumers and commodities, respectively. Based on this model, a distributed user association method is developed that also takes advantage of the Walras' tatonnement process. As another example, in \cite{Maghsudi16:DUA}, downlink user association is performed in a distributed manner at end-users, where every user aims at maximizing the probability of successful transmission. In both works, every user can be associated to multiple SBSs, which simplifies the handover process.
\subsubsection{Energy efficiency}
\label{subsubsec:EEff}
It is known that in UD-SCNs, energy efficiency and capacity are improved when small cells are turned on and off using a sleep-awake scheduling method \cite{Chang14:CME}, called \textit{cell planning} or \textit{cell scheduling}. It is however clear that turning SBSs on and off has to be performed in combination with a sophisticated user association mechanism, in a way that changing SBS density does not degrade the user satisfaction level. In \cite{Dong14:EEP}, the authors optimize the SBS density for energy efficiency in cellular networks by using stochastic geometry, and optimize the user association matrix by using quantum particle swarm optimization. Similarly, in \cite{Kim13:AJA}, two heuristic algorithms are proposed to jointly optimize base station operation and user association in heterogeneous networks. However, none of these works considers energy harvesting in the system model.  
\begin{figure}[t]
\centering
\includegraphics[width=0.49\textwidth]{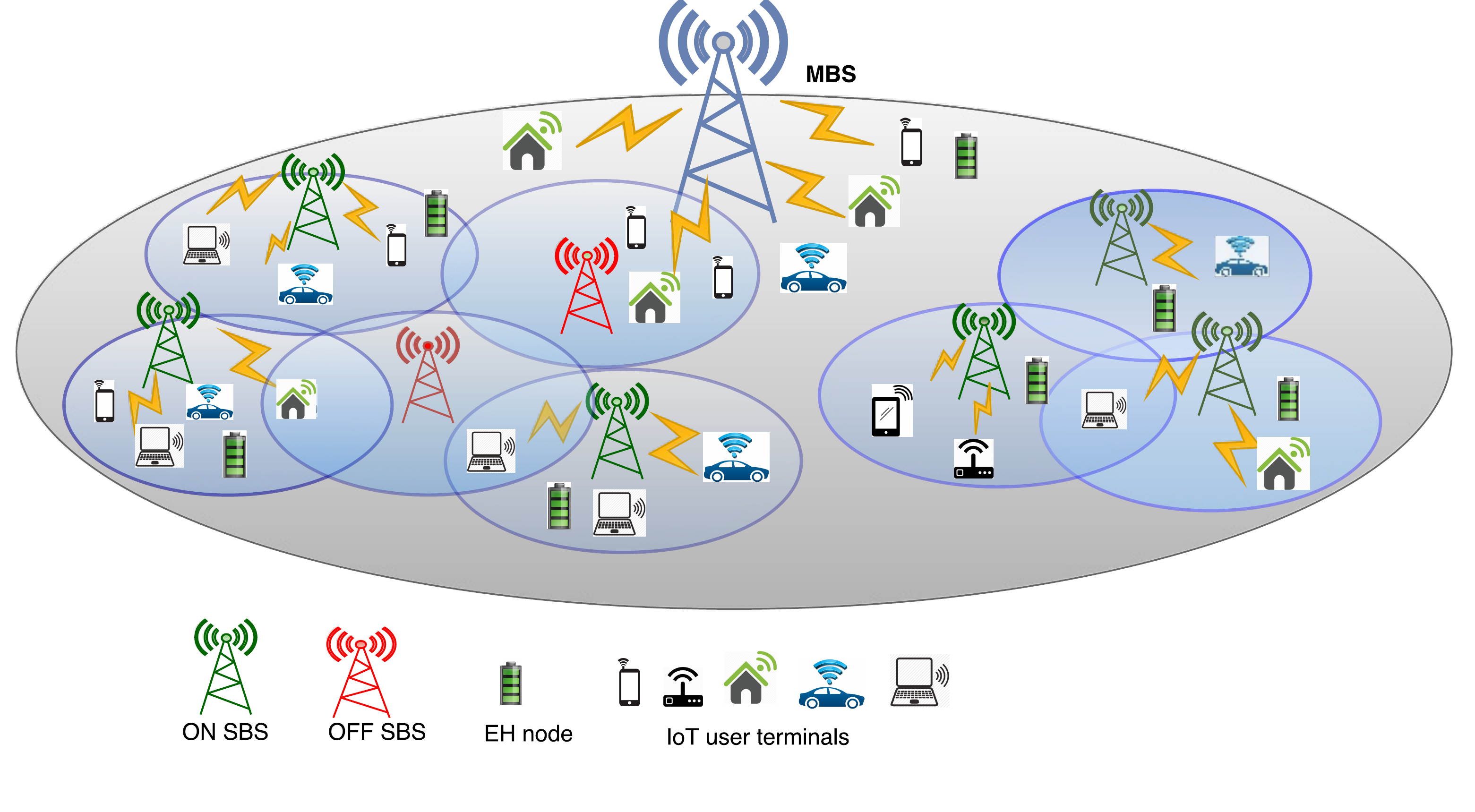}
\caption{Cell planning and user association in IoT-driven UD-SCNs with energy harvesting.}
\label{Fig:UD-SCN}
\end{figure}

A short summary of cell association schemes in SCNs is provided in Table \ref{Tb:SoA}. Note that the table is not exhaustive since the number of references should be kept limited.
\begin{table*}
\scriptsize
\caption{Cell association in SCNs} 
\label{Tb:SoA}
\centering 
\begin{tabular}
{|c|c|c|c|c|}
\hline 
Reference & Objective & Method & Energy harvesting & Limitation\\ 
\hline
\hline
\cite{Maghsudi15:DUA} & Aggregate throughput maximization & Exchange economy with uncertainty & Yes & Control channel or data exchange \\ 
\hline
\cite{Maghsudi16:DUA} & Maximize transmission success probability & Multi-armed bandits & Yes & Slow convergence\\
\hline
\cite{Dong14:EEP} & Energy-efficient cell planning & Stochastic geometry, particle swarm optimization & No & Information acquisition, computation \\ 
\hline
\cite{Kim13:AJA} & Energy-efficient cell planning & Combinatorial optimization & No & Information acquisition \\ 
\hline
\cite{Li:SIC} & Interference mitigation & Stochastic geometry & No & Information acquisition, computation \\ 
\hline
\cite{Chiang:JPC} & Interference mitigation & Integer programming & No & Information acquisition, computation \\ 
\hline
\end{tabular}
\end{table*}
\subsection{Open Issues}
\label{subsec:open}
Despite being under intensive investigation, many features of IoT-driven ultra-dense SCNs remain unexplored. In particular, enhancing the network operation by energy harvesting impacts a large body of conventional problems dramatically, so that the solution approaches developed so far are no longer useful. For example, although cell planning in conjunction with user association has been investigated before (see Section \ref{subsubsec:EEff}), enabling SBSs and/or users to harvest the ambient energy renders the existing solutions non-applicable, thereby necessitating the development of new solution approaches. In the following, we discuss some open problems.  
\subsubsection{Distributed cell association for massive IoT devices}
\label{subsec:FRD}
As described before, cell association is a fundamental problem which affects the performance of SCNs in a variety of aspects. For instance, the quality of service experienced by every user depends strongly on association. Moreover, most emerging networking concept, for instance caching and full-duplexing, can be efficiently implemented only in conjunction with intelligent cell association. Although a large body of literature propose centralized approaches to solve the cell association problem, such methods require costly information acquisition and heavy computations, thus are not practical for ultra-dense small cell networks. Moreover, a great majority of previous studies neglect the energy efficiency problem by excluding the randomness of energy harvesting from the analysis. Thus, for energy-harvesting UD-SCNs, it is beneficial to formulate the (joint) uplink and downlink cell association as distributed optimization problem under uncertainty. In most cases, the formulated optimization problem can be afterward modeled as distributed decision making problem is a large multi-agent system with uncertainty. However, naive learning methods such as regret matching or conventional solution concepts such as Nash equilibrium are not sufficient to deal with (very) large and non-deterministic systems. Such solutions mostly i) require unaffordable prior knowledge at every agent, ii) are not able to deal with the uncertainty, iii) converge slowly for large number of actions and agents. In Section \ref{sec:MFBG}, we will describe an efficient mathematical model to analyze UD-SCNs that does not suffer from such shortcomings. Afterward, in Section 
\ref{sec:EH}, we solve the uplink distributed cell association problem using this solution approach.
\subsubsection{Uplink traffic reduction and interference management}
\label{subsubsec:Uplink}
Conventionally, cellular systems are designed to perform the transmission of large data streams in the downlink, in order to serve human-driven service demands. Nevertheless, by emerging networking concepts such as IoT, the next generation wireless networks are required to also support a variety of machine-type communications, for instance information transmission in wireless sensor networks used for site surveillance \cite{Shariat15:MAC}. Such data streams are small individually but create a heavy traffic in the uplink when they superimpose. Moreover, closely located users that transmit in the same frequency band cause interference to each other. Changing traffic pattern and mitigating the interference can be realized by intelligent user association. 
\subsubsection{Transmission mode selection}
\label{subsubsec:Mode}
In heterogeneous wireless networks, nearby users might be able to establish direct communication links instead of transmitting the data via an SBS or the MBS. In comparison with multi-hop communications via an access point, direct communications is shown to be more efficient (in terms of power and spectrum) and secure, especially for cell edge users; however, it changes the network traffic and interference pattern dramatically. Moreover, enhancing the network with the possibility of direct transmission, gives rise to the problem of \textit{transmission mode selection}. In order to reduce the overhead of information acquisition and computational cost, transmission mode selection has to be performed by users in a distributed manner. In addition, in UD-SCNs where every user might be in the coverage area of multiple SBSs, mode selection has to be performed jointly with cell association, since the transmission performance varies across SBSs. As described before, the uncertainty of energy harvesting and lack of channel knowledge make the problem more complicated, since the (a priori unknown) available energy as well as user's channel matrix play an important role in selecting the optimal transmission mode and (possibly) consequent cell association. In \cite{Maghsudi14:TMS}, the authors propose a distributed mode selection method in heterogeneous networks by using multi-armed bandit theory, where users have only very limited information; nonetheless, energy harvesting is not included in the system model.  
\subsubsection{Full-duplex transmission}
\label{subsubsec:Full}
Full-duplex transmission\footnote{Simultaneous transmission and reception in the same frequency band is referred to as 
in-band full-duplexing and simultaneous transmission and reception in different frequency bands is referred to as 
out-of-band full-duplexing.}~is an emerging technology which has a potential to significantly increase the spectral efficiency. Recent studies show that in-band full-duplex technology exhibits better performance for low power transmission nodes, which makes it suitable for SCNs. Nonetheless, in such systems, it is inefficient to perform the uplink and downlink user association separately, since uplink and downlink transmissions are mutually dependent due to self-interference. More precisely, it should be noted that an efficient user association in the downlink 
(uplink) might cause high interference in the uplink (downlink). Taking the uncertainty of energy harvesting into consideration makes the problem more difficult, since in this case the set of users that can be successfully served by an SBS (in uplink and downlink) is non-deterministic. While this problem is already challenging to be solved by a central controller, it becomes much more difficult if all network nodes take active roles in selecting their correspondence. In this case, it is vital to search for decision making methods that converge to an efficient equilibrium or steady state. While user association in full-duplex SCNs has been addressed previously (see \cite{Sekander15:AMG} for an example), investigating the effect of energy harvesting is an open issue.  
\subsubsection{Caching and multicasting}
\label{subsubsec:Cash}
In the past few years, caching has attracted an ever-increasing attention from the communications society as a potential solution to reduce the backhaul traffic and to gain spectral efficiency. The basic idea of femto-caching\footnote{Along with femto-caching, another important strategy is D2D-caching, where clustered devices build up a virtual library through storing different files privately and sharing them whenever necessary.}~is to save popular files (i.e., those files which tend to be requested by many users in short time intervals) in cache-enabled SBSs, instead of fetching the data frequently from the core network. In a densely deployed SCN, every small cell has a small coverage area and serves mainly its nearby users. Meanwhile, it is noticeable that normally, popular contents are requested by multiple users simultaneously. Thus, a feasible solution to reduce the SBS traffic, also to offload the wired backhaul, would be caching popular context and then to employ multicasting for data transmission. This concept itself gives rise to context-aware user association, where users are clustered and associated based on the possibility of demanding similar files, which can be predicted using machine learning methods and historical requested data. A similar concept is discussed in \cite{Poularakis14:MAC}, where the MBS multicasts files to cache-enabled SBSs. All interested SBSs are then able to cache the file, instead of fetching it individually from the core network.  
\section{Mean-field Bandit Games}
\label{sec:MFBG}
\subsection{Single-agent Multi-armed Bandits}
\label{subsec:SA-MAB}
Multi-armed bandits (MAB) can be categorized as a class of sequential optimization problems, where given a set of arms 
(actions), a player pulls an arm at successive trials to receive some a priori unknown reward. Upon pulling an arm, the player observes only the reward of the played arm, and not those of other arms. Due to the lack of information, there might be some difference between the maximum reward (achievable by pulling the optimal arm), and the reward of the actual played arm. This difference is generally referred to as \textit{regret}. The player decides which arm to pull in a sequence of trials so as its average regret is minimized, or its (discounted) average reward is maximized. The basic problem here is to deal with the famous exploration-exploitation dilemma, i.e., to find a balance between receiving immediate rewards (exploitation) and gathering information to achieve large reward only in the future (exploration). 

Different types of bandit games are defined based on i) the random nature of reward generating processes, ii) density and type of agents, iii) availability of side-information, iv) randomness in action availability, and so on. To date, a variety of algorithmic solutions (policy) are developed to solve different types of single-agent bandit problems. A 
well-known policy, which we also use later in this paper, is the upper confidence bound (UCB) algorithm. The seminal UCB algorithm is designed specifically for stochastic stationary bandit problems, where the set of instantaneous rewards of each arm are independent and identically-distributed (i.i.d) random variables. At every round of selection, the UCB algorithm estimates an upper-bound of the mean reward of each arm at some fixed confidence level. The arm with the highest estimated bound is then played, and bounds are updated after observing the reward. For a survey on the types and solutions of bandit problems, see \cite{Maghsudi15:MSC}. 
\subsection{Multi-agent Multi-armed Bandits}
\label{subsec:MA-MAB}
When multiple agents are involved in a bandit game, the agents affect each other in the sense that the reward achieved by every agent is determined not only through its own actions, but also through the joint action profile of other agents. In other words, the payoff of every arm to every agent depends not only on the type (or ability) of that specific agent, but also on the number of agents selecting that arm. For instance, in a congestion model, the individual rewards might decrease in case multiple agents select an arm, whereas in a coordination model, the reward might increase. In addition to minimizing the regret, in a multi-agent setting, it is important to reach some sort of system stability or equilibrium. 

For small number of agents, there are some results that connect multi-agent multi-armed bandit games with correlated and Nash equilibria (see \cite{Maghsudi15:MSC} for a review). Perfect Bayesian equilibrium is another notion of equilibrium that is widely-used in conjunction with learning games. For multi-armed bandit games with large number of agents, however, such equilibrium notions are not practical since they yield excessive complexity and long convergence time. For example, for a multi-armed bandit game to converge to correlated equilibrium, every agent has to observe the joint action profile of other agents (full monitoring) and forecast their future moves, which is clearly a highly-involved task even when moderate number of agents compete with each other. 
\subsection{Mean-field Model for Multi-agent Multi-armed Bandits}
\label{seubsec:MFM}
In general, games with (very) large number of agents are analyzed using mean-field models. In mean-field models, every agent regards the rest of the world as being stationary, considering individual moves of agents as unimportant details. In other words, mean-field analysis restates game theory as an interaction of each individual with the mass of others. While mean-field analysis for games with perfect information is well-established, applying this concept to multi-armed bandit games is a recently-emerging research direction. In \cite{Gummadi12:MFE}, Gummadi et al. provide basic elements of such models. In what follows, we describe mean-field multi-armed bandit games briefly, without getting into mathematical details. To show an application of this model, in Section \ref{sec:EH}, we will return to energy-harvesting UD-SCNs and model the uplink cell assignment problem by mean-field multi-armed bandit games.  

Consider a multi-agent multi-armed bandit game $\mathfrak{G}$ that consists of a set $\mathcal{M}$ of $M$ arms and a set 
$\mathcal{N}$ of $N$ agents. At every round $t$, each agent $n \in \mathcal{N}$ selects an action, denoted by $a_{n,t}$, from the predefined action set $\mathcal{M}$ and receives some a priori unknown reward. In the mean-field model of game 
$\mathfrak{G}$, each agent $n \in \mathcal{N}$ is characterized by some type $\theta_{n,t} \in \left[0,1 \right]^{M}$ and some state $\mathbf{Z}_{n,t}$. Let $f_{m,t}$ denote the fraction of agents that pull arm $m \in \mathcal{M}$ at trial $t$. Then the reward of arm $m \in \mathcal{M}$ to an agent $n \in \mathcal{N}$ is a Bernoulli random variable with some parameter $Q(f_{m,t},\theta_{n,t})$. The state of each agent is simply its game history, i.e., its past actions and rewards. More formally, $\mathbf{Z}_{n,t}=\left(w_{1,t},l_{1,t},...,w_{M,t},l_{M,t} \right)$, with $w_{m,t}$ and $l_{m,t}$, $m \in \mathcal{M}$, being the total number of successes and failures of arm $m$ up to time $t$ when selected by agent 
$n \in \mathcal{N}$. As it is conventionally assumed in mean-field game models, every agent regenerates after some random time which follows a geometric distribution with parameter $1-\alpha$, $\alpha \in [0,1)$. Note that in the literature of mean-field games, regeneration means that an agent quits the game and a new agent takes its place.

For an arbitrary agent $n \in \mathcal{N}$, the \textit{mean-field dynamics} works as follows: If a trial is a regeneration trial (by chance), the agent's type $\theta_{n,t}$ is sampled from some distribution $W$, where samples are i.i.d. Moreover, the state $\mathbf{Z}_{n,t}$ is reset to zero. Otherwise (in case of no regeneration), the type remains unchanged, and a (randomized) selection policy (for example, the UCB algorithm) is used to map $\mathbf{Z}_{n,t-1}$ to some action $a_{n,t}$. It is assumed that all agents use the same policy $\delta$ throughout the game. The agent then receives some random reward following a Bernoulli distribution with parameter $Q(f_{m,t},\theta_{n,t})$ and the state vector $\mathbf{Z}_{n,t}$ is updated. Details can be found in \cite{Gummadi12:MFE}. Also, the following results on the existence and uniqueness of 
mean-field equilibrium are stated in \cite{Gummadi12:MFE}, \cite{JohariSlides}. 
\begin{theorem}
\label{Pr:Cont}
If $Q$ is continuous in $f$, then a mean-field equilibrium exists.
\end{theorem}
\begin{theorem}
\label{Th:Lips}
Suppose that $Q$ is Lipschitz in $f$ with Lipschitz constant $L$. Then if $\alpha (1+L)<1$, then there exists a unique 
mean-field equilibrium, and the mean-field dynamics converges to it from any initial condition. 
\end{theorem}

To date, the mean-field multi-armed bandit model is studied only for reward generating process following Bernoulli distribution; this assumption however restricts the applicability of the model. Thus a future line of research might be to generalize the analysis (including mean-field dynamics, existence and uniqueness of mean-field equilibrium) to other models of rewards' randomness. 
\section{Uplink Cell Association for Energy Harvesting IoT Devices in Small Cell Networks}
\label{sec:EH}
As described before, although a rich body of literature is dedicated to investigate the cell association problem, most of the existing methods i) are developed for downlink transmission, ii) rely on a central controller, and/or iii) demand excessive amount of information at SBSs or user devices. Thus, desired is to develop a distributed association method which is capable to deal with very large network size (number of SBSs and/or users) as well as information shortage. Moreover, within the scope of energy harvesting networks, an efficient solution to this problem ought to address the uncertainty imposed by random energy arrivals and/or intensities.  
\subsection{System Model and Assumptions}
\label{subsec:SysModel}
We consider a dense small cell network that consists of a set $\mathcal{M}$ of $M$ small cells and a set $\mathcal{N}$ of 
$N$ devices. Every device $n \in \mathcal{N}$ intends to transmit $J_{n} \leq J$ data packets in the uplink direction in successive transmission rounds.\footnote{Note that when a device quits transmission, it is replaced by another device so that the number of devices is always equal to $N$. As described before, this corresponds to 
\textit{regeneration} in mean-field game models and is assumed to follow a geometric distribution.}~At every transmission round $j$, each device transmits one data packet to an SBS of its choice, implying that the association is performed in a distributed manner. Note that in what follows we omit the time notion $j$ for the simplicity of notation unless an ambiguity arises. Multiple devices can be served by a single SBS. By $\mathcal{N}_{m,j}$ we denote the set of $N_{m,j}$ devices to be served by SBS $m \in \mathcal{M}$ at round $j$. Every device obtains the energy through ambient energy harvesting by applying a \textit{harvest-store-use} strategy; that is, for every transmission round, it harvests the energy and then uses \textit{all} of it for transmission. Energy harvesting is independent across devices. Without loss of generality, we assume that the energy equals power. 

Since energy harvesting is opportunistic, for every device $n$ and at every round $j$, the amount of harvested energy, denoted by $P_{n,j}$, is unknown a priori. We assume that $P_{n,j}$, $j=1,...,J_{n}$, are i.i.d random variables following half-normal distribution (a special case of folded normal distribution) with parameter $\sigma_{n}^{2}>0$. This assumption is not restrictive since the half-normal distribution can be replaced by any other distribution, without affecting the solution approach.  

The $N_{m,j}$ devices which select any SBS $m \in \mathcal{M}$ share the available spectrum resources equally in an orthogonal manner; that is, inside every small cell, transmissions are corrupted only by zero-mean additive white Gaussian noise with variance $N_{0}$. For each small cell $m \in \mathcal{M}$, the inter-cell interference experienced by every device $n \in \mathcal{N}_{m,j}$, denoted by $I_{nm} \geq 0$, is regarded as noise and assumed to be fixed during the entire transmission. At round $j$, the real-valued channel gain between device $n \in \mathcal{N}_{m,j}$ and small cell 
$m \in \mathcal{M}$ is denoted by $h_{nm,j}$. We assume frequency non-selective block fading channel model, where the random variable $H_{nm}$ follows a Rayleigh distribution with parameter $\frac{1}{\sqrt{2\beta_{nm}}}$, and remains constant during the transmission of every packet $j=1,...,J_{n}$ for all $n \in \mathcal{N}$ and $m \in \mathcal{M}$, and changes from one transmission round to another. The random channel gain $H'_{nm}=H_{nm}^{2}$ then follows an exponential distribution with parameter $\beta_{nm}$. At every round $j$, the \textit{type} of every device $n \in \mathcal{N}$ is defined as the collection of its channel gains towards SBSs, i.e., $\mathbf{h}'_{n,j}=\left(h'_{n1,j}, h'_{n2,j},..., 
h'_{nM,j} \right)$, $\mathbf{h}'_{n,j} \in (0,1]^{M}$.\footnote{Note that assuming 
$I_{nm}$ to be time-invariant only simplifies the notation and does not result in a loss of generality, due to the following reason: If the interference is time-varying, so denoted as $I_{nm,j}$, the type can be simply defined as the collection of channel gain to interference-plus-noise ratio of a user with respect to all small cells. Formally, let 
$\theta_{n,t}$ be the type of user $n \in \mathcal{N}$. Then by the new definition of type we have $\theta_{n,t}=
\left(h'_{n1,j}/(I_{n1,j}+N_{0}), h'_{n2,j}/(I_{n2,j}+N_{0}),..., h'_{nM,j}/(I_{nM,j}+N_{0}) \right)$. If $I_{nm,j}$ is deterministic, the {\em type} still follows the same distribution as $h'_{nm}$ but with different parameters. If 
$I_{nm,j}$ is a random variable, the distribution of type can be calculated by using simple rules of probability. In both cases, the model is still applicable.}~Note that the devices do not have any channel state information. In other words, the {\em type} is unknown. Let $f_{m,j}= \frac{N_{m,j}}{N}$ denote the fraction of devices that select SBS $m$ at round 
$j$. Thus, for each $n \in \mathcal{N}_{m,j}$ and for transmitting every data packet $j$, the achievable transmission rate is given by 
\begin{equation}
\label{eq:Rate}
r_{nm,j}=\frac{W_{m}}{Nf_{m,j}} \log \left(1+\frac{P_{n,j}h'_{nm,j}}{I_{nm}+N_{0}} \right),
\end{equation}
where $W_{m}$ is the available bandwidth at SBS $m \in \mathcal{M}$. Moreover, as stated before, $P_{n,j}$ is the transmit power of device $n$ at trial $j$, which equals the amount of energy harvested at that trial. For transmission of every data packet, every device $n \in \mathcal{N}$ requires a specific quality of service (QoS) that is expressed in terms of a minimum data rate $r_{n,\min}$. Hence, for any device $n \in \mathcal{N}$, at every transmission round $j$, we define the reward of selecting SBS $m \in \mathcal{M}$ as 
\begin{equation*}
\label{eq:Utility}
u_{n,j}(m)=\begin{cases}
1, & \textup{if} \hspace{5pt}r_{nm,j}\geq r_{n,\min}, \\ 
0, & \textup{otherwise}. 
\end{cases}
\end{equation*}
The success probability of user $n$ when selecting SBS $m$ at every transmission round $j$ is then given as 
\begin{equation*}
\label{eq:success}
p_{nm,j}^{(s)}=\textup{Pr}\left[r_{nm,j} \geq r_{n,\min} \right],
\end{equation*}
and the failure probability yields $p_{nm,j}^{(f)}=1-p_{nm,j}^{(s)}$. Thus, successful transmission is a Bernoulli random variable with parameter $p_{nm}^{(s)}$.

From (\ref{eq:Rate}), it can be easily concluded that given $r_{n,\min}$, for any specific $f_{m,j}$ and $h'_{nm,j}$, 
\begin{equation}
\label{eq:minPower}
P_{n,j,\min}= \frac{I_{nm}+N_{0} }{h'_{nm,j}}\left (e^{ \frac{Nf_{m,j}r_{n,\min}}{W_{m}}}-1  \right )
\end{equation}
yields $r_{nm,j} \geq r_{n,\min}$. Thus we have  
\begin{equation*}
\label{eq:successT}
p_{nm,j}^{(s)}=\textup{Pr}\left[P_{n,j} \geq P_{n,j,\min} \right].
\end{equation*}
By the half-normal assumption on the harvested energy we can conclude 
\begin{equation}
\label{eq:successTr}
p_{nm,j}^{(s)}=1-\textup{erf} \left[\frac{P_{n,j,\min}}{\sqrt{2}\sigma_{n}}\right],
\end{equation}
so that roughly, $p_{nm,j}^{(s)}\propto \frac{h'_{nm,j}}{f_{m,j}}$, which, as expected, corresponds to a congestion model. 
\subsection{Mean-field Equilibrium of the Multi-armed Bandit Game Model}
\label{subsec:MFE}
According to the system model described before, the {\em type} of each device $n \in \mathcal{N}$ is the collection of channel gains towards SBSs. Upon selecting any SBS $m \in \mathcal{M}$ at round $j$, the device transmits successfully (receives reward) with probability $p_{nm,j}^{(s)}$. The success probability depends on the {\em type} of user $n$ as well as the fraction of devices that select SBS $m$, $f_{m,j}$. Thus, the uplink cell association can be cast as a multi-armed bandit game. Assuming that the number of devices, $N$, is large, the following proposition describes the characteristics of mean-field equilibrium in multi-armed bandit game model of the formulated uplink cell association problem.
\begin{proposition}
In the mean-field multi-armed bandit game model for the uplink cell association problem, the following results hold.\footnote{Note that here we omit the time subscript $j$ to simplify the notations.}
\begin{enumerate}
\item There exists a mean-field equilibrium.
\item Let $a_{nm}=\sqrt{\frac{2}{\pi}}\frac{Nr_{n,\min}(I_{nm}+N_{0})}{W_{m}h'_{nm}\sigma_{n}}$ and $b_{nm}=
\frac{\left (I_{nm}+N_{0} \right)^{2}}{2h'_{nm}\sigma_{n}^{2}}$. Moreover, assume that for all $n \in \mathcal{N}$ 
and $m \in \mathcal{M}$, the regeneration parameter $\alpha$ (see Section \ref{sec:MFBG}) satisfies 
$\alpha \leq \frac{1}{1+a_{nm}e^{b_{nm}}}$. Then the mean-field equilibrium is unique and the mean-field dynamics converges to it from any initial point.
\end{enumerate}
\end{proposition}
\begin{IEEEproof}
The first part follows immediately from Theorem \ref{Pr:Cont}. The second part follows by some simple calculus steps as sketched in the following: We substitute (\ref{eq:minPower}) in (\ref{eq:successTr}), then calculate 
$\frac{\partial p_{nm}^{(s)}}{\partial f_{m}}$. Since $0\leq f_{m}\leq 1$, we have 
$\frac{\partial p_{nm}^{(s)}}{\partial f_{m}} \leq a_{nm}+e^{b_{nm}}$. Thus the result follows by Theorem \ref{Th:Lips}. 
\end{IEEEproof}
\subsection{Performance Evaluation Results}
\label{subsec:Per}
We consider a small cell network where the devices apply the mean-field dynamics described in Section \ref{sec:MFBG}, with the policy $\delta$ being the UCB algorithm. The goal of every device is to successively decide for the SBS which, with the highest probability, yields a reward; that is, the arm with the largest success probability. Without loss of generality and for simplification, we choose $W_{m}=N$, $r_{n,\min}=0.75$, $I_{nm}+N_{0}=1$, and $\sigma_{n}=1$, for all 
$n \in \mathcal{N}$ and $m \in \mathcal{M}$. For every device $n \in \mathcal{N}$, the collection of channel gains 
$\mathbf{h}'_{n}$ (i.e., the {\em type}) is selected at random for initialization and upon regeneration. Note that for every device $n \in \mathcal{N}$, and for every SBS $m \in \mathcal{M}$, $h'_{nm}$ are independent but non-identically distributed. The state $\mathbf{Z}_{n}$ is initialized randomly.  

\begin{figure}[t]
\centering
\includegraphics[width=0.49\textwidth]{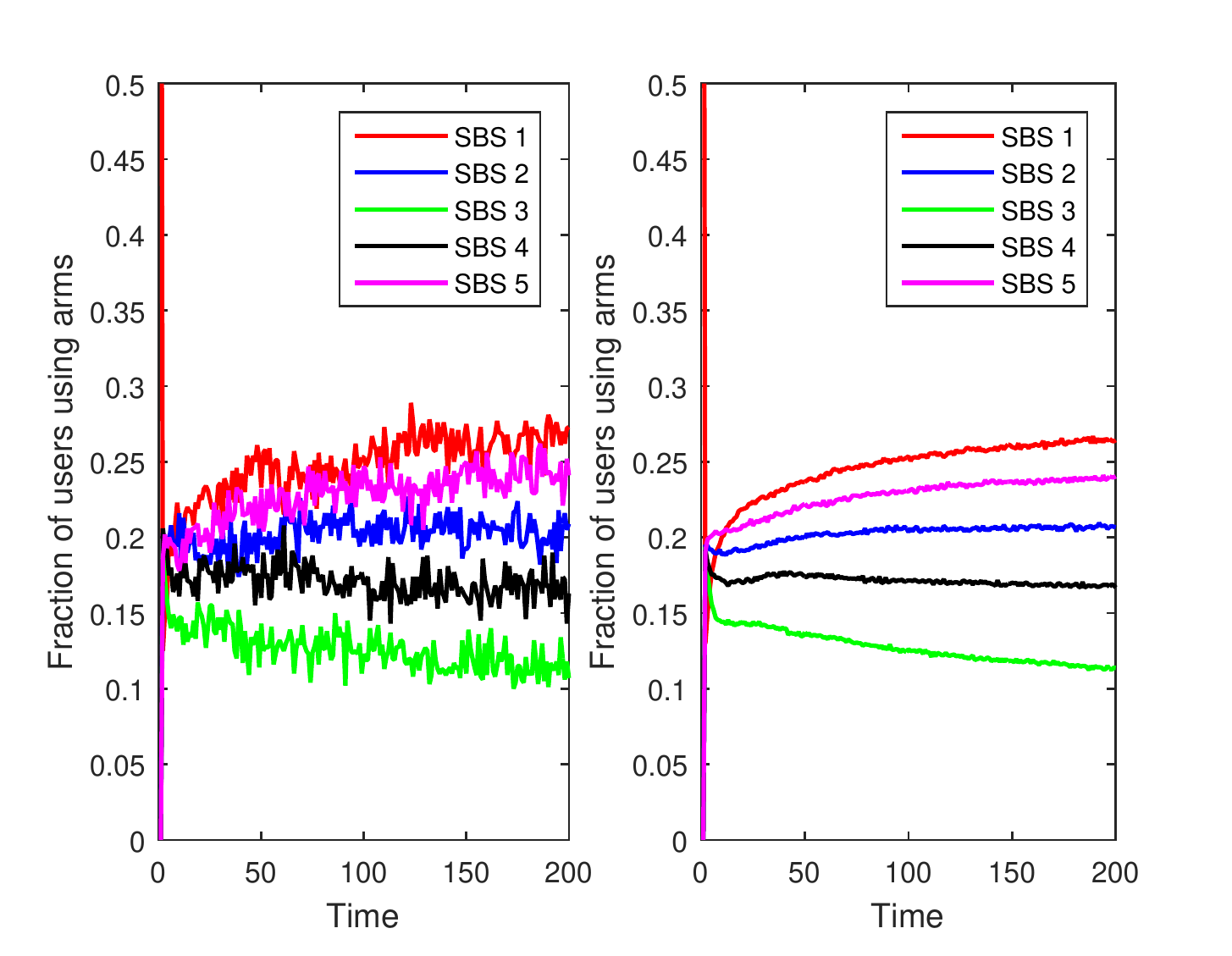}
\caption{The effect of the number of devices on the performance of mean-field dynamics (left: $10^3$ devices, 
right: $5 \times 10^4$ devices).}
\label{Fig:UserEff}
\end{figure}

Fig. \ref{Fig:UserEff} shows the effect of the number of devices on the performance of mean-field multi-armed bandit dynamics when the number of SBSs is selected as $M=5$. As expected, the dynamics performs well in particular for very large number of devices. The reason is as follows: In mean-field dynamics, every device considers the rest of the world as stationary stochastic, ignoring individual moves by considering them as unimportant details. This assumption is specifically valid for large number of devices, since in a small system individual moves might have great impact and thus cannot be neglected.

\begin{figure}[t]
\centering
\includegraphics[width=0.49\textwidth]{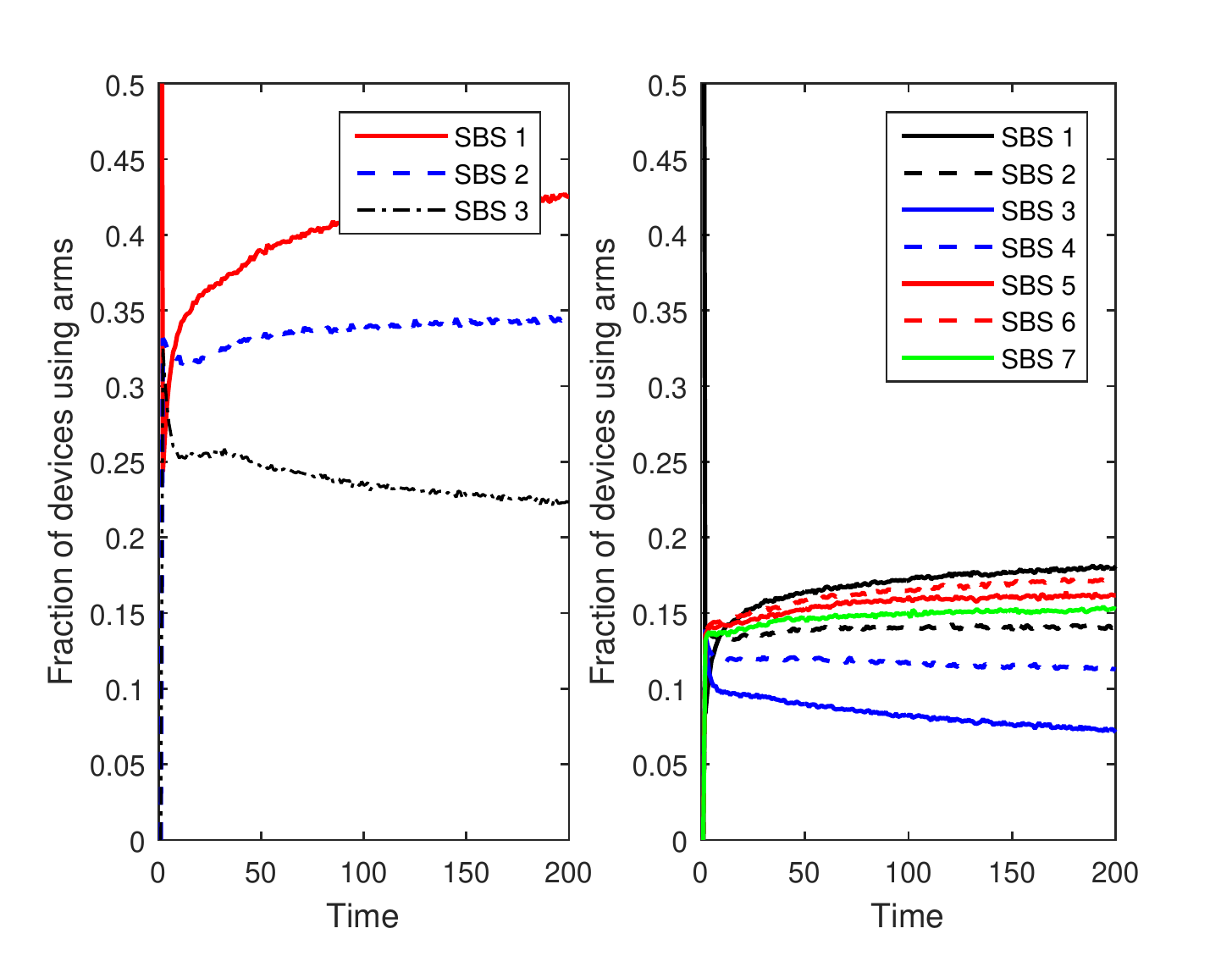}
\caption{The effect of the number of SBSs on the performance of mean-field dynamics (left: 3 SBSs, right: 7 SBSs).}
\label{Fig:ActEff}
\end{figure}

In Fig. \ref{Fig:ActEff}, we select the number of devices $N=5 \times 10^{4}$ and investigate the effect of the number of SBSs (arms) on the performance of mean-field multi-armed bandit model. The figure shows the convergence of the algorithm regardless of the number of devices.
%
 
\begin{figure}[t]
\centering
\includegraphics[width=0.45\textwidth]{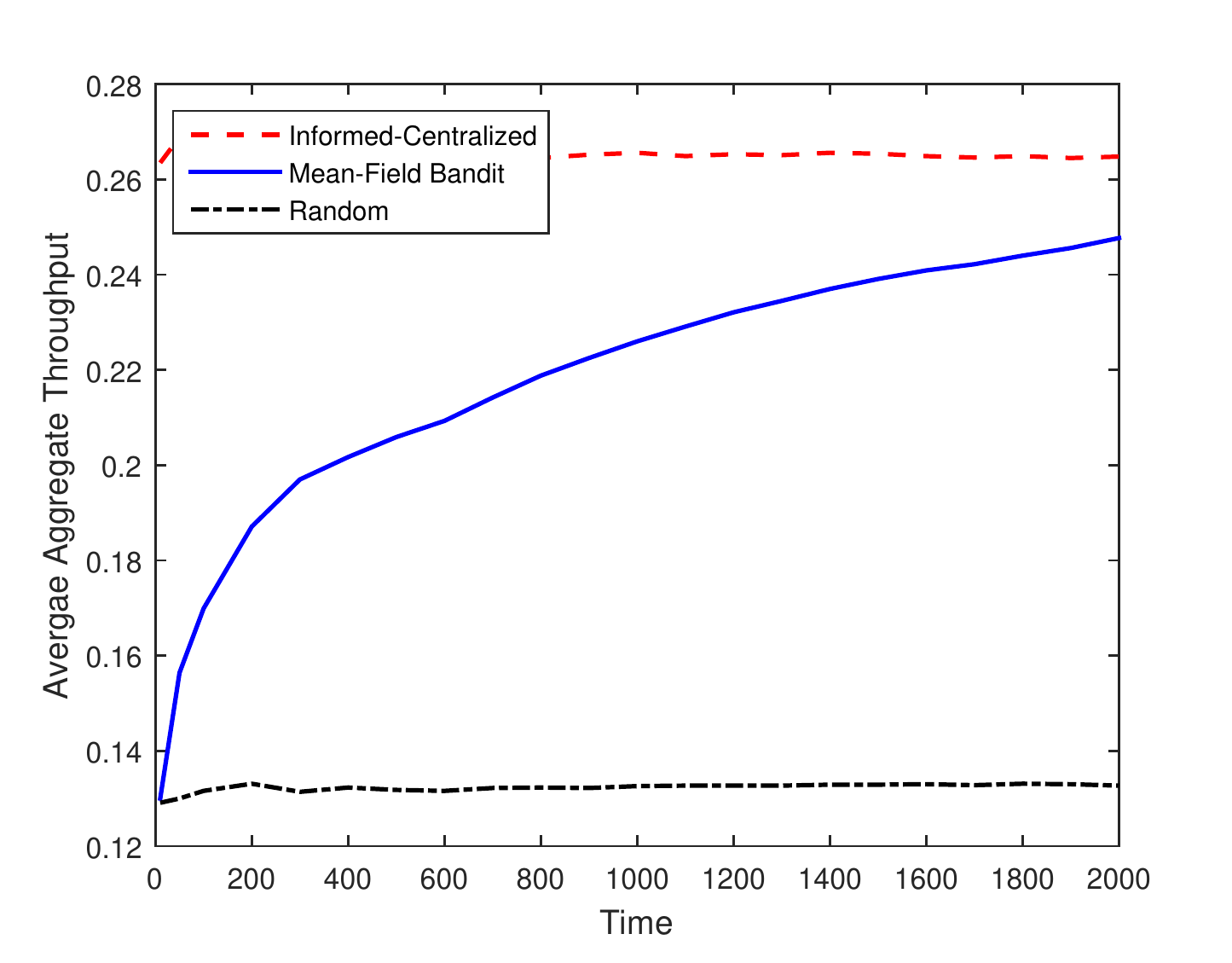}
\caption{The performance of mean-field bandit model compared with Centralized assignment given type information and random selection.}
\label{Fig:Comp}
\end{figure}

In Fig. \ref{Fig:Comp}, we select the number of devices $N=10^{3}$ and the number of SBSs $M=3$. We compare the performance of mean-field multi-armed bandit, in terms of aggregate average throughput, with those of following assignment techniques:
\begin{itemize}
\item {\em Optimal (centralized-informed) assignment}: In this scenario, cell association is performed by a central unit given types, $\mathbf{h}'_{n}$, for all $n \in \mathcal{N}$. By means of exhaustive search, every device is assigned to the SBS to which it has the maximum average channel gain. 
\item {\em Random assignment}: Every device selects an SBS simply at random.
\end{itemize} 
%
The figure shows that the mean-field bandit performs well. Note that despite its slightly better performance, centralized-informed assignment yield excessive complexity and overhead and thus is impracticable in particular for networks with very large number of devices.

The time and space complexity of the mean-field dynamics depend on the policy $\delta$ (see Section \ref{sec:MFBG}), which maps the state to the action in case no regeneration takes place. As mentioned before, here we use the UCB algorithm which calculates an index for each action corresponding to some confidence bound. The arm with the largest index is then played. Thus, the time and space complexity is polynomial in the number of actions. Moreover, note that the notion of \textit{optimality} is not part of the definition since the policy $\delta$ is fixed a priori \cite{Gummadi12:MFE}. In fact, the most important metric to be observed is the convergence to mean-field equilibrium. We should also emphasize that the algorithm can be implemented fully distributively and no other overhead/complexity incurs. In addition, it is worth noting that our work is the first cell association model that combines the uncertainty of (both) energy harvesting and wireless channel with the high density of SBSs and devices in SCNs using the mean-field multi-armed bandit model. Thus, the performance of no other cell association method can be fairly compared with ours. 

For future research, the system model might be improved to include intra-cell interferences, where in every cell users cause interference to each other. Another possibility is to investigate a cell association problem in full-duplex SCNs, where uplink and downlink transmissions are mutually dependent due to self-interference.  
\section{Conclusion}
\label{sec:Conclusion}
We have studied a wireless networking paradigm for IoT, namely, small cell networks, where densely-deployed small base stations serve a large number of devices (human-driven devices and/or machines). We have assumed that the energy by the IoT devices is obtained through local ambient energy harvesting. Due to its opportunistic nature, this, in addition to the random wireless channel, introduces some uncertainty in the network operation. We have focused on the distributed cell association problem in such networks and discussed the state-of-the-art as well as challenges and future research directions. We have described a mean-field multi-armed bandit game model, in which a large number of agents with limited information, sequentially select an action from a finite action set, thereby affecting the welfare of each other. Due to the mean-field analysis, this model is particularly appropriate for analyzing the devices' interactions in dense small cell networks with strictly-limited prior information. With energy harvesting at the IoT devices, we have modeled the uplink cell association problem by a mean-field multi-armed bandit game. We have showed that a unique mean-field equilibrium exists to which the mean-field dynamics converges from any initial condition. We have also provided some numerical results which show the applicability of the model to networks with very large number of devices.  
\bibliographystyle{IEEEbib}
\bibliography{main}
\end{document}